# Optical Lossy-mode-resonance Relative Humidity Sensor on a Fiber Tip

Yundong Ren[1], Mucheng Li[1], and Yuxiang Liu[1,*]

[1]Department of Mechanical Engineering, Worcester Polytechnic Institute, Worcester, Massachusetts 01609, USA

*Corresponding author. Email: yliu11@wpi.edu

**Abstract**

Real-time measurement of relative humidity (RH) is important to many physical, chemical, and biological processes. However, in processes that involve harsh conditions such as high temperature, strong electromagnetic interferences, and complex spatial constraints, conventional electrical sensors often fall short due to their intrinsic limitations. In this work, we developed an optical lossy-mode-resonance (LMR) RH sensor based on the $SnO_2$ film coated D-shaped fiber tip. Thanks to the high-temperature endurance and electromagnetic interference immunity, the developed optical fiber-tip sensor is ideal for RH sensing in a critical environment, such as in the microwave drying and hot air drying processes. Furthermore, unlike other reported LMR sensor with an in-line form factor, our sensor is located at the fiber tip with a probe-like form factor, allowing it to be readily implemented in a spatially confined environment. We have developed a custom setup that allows the fabrication of the novel D-shaped LMR fiber-tip sensor. The LMR signal from the D-shaped fiber-tip sensor was experimentally characterized. The lossy mode resonances are understood by the finite element analysis, the results of which agree well with the experimental measurements. The fiber-tip sensor had a linear RH response between 6.1% to 75.0% and a resolution better than 4.0% RH. The fiber-tip sensor demonstrated a response time and reversibility comparable to that of the commercial electrical sensors in the RH range of ~5.0% to ~86.0%. The novel D-shaped fiber-tip LMR RH sensor developed in this work will benefit many applications that require RH monitoring in a harsh critical environment. Furthermore, our innovative D-shaped fiber tip design can be readily applied to LMR-based fiber sensors in general. This will allow the design's advantages of small footprint and agile maneuverability to benefit a wide range of LMR fiber sensing applications beyond RH sensing.

.

## I. Introduction

Real-time relative humidity (RH) plays a critical role in many physical, chemical, and biological processes [1]–[4]. As a result, real-time environmental RH monitoring is of great importance in a wide range of applications, including but not limited to food manufacturing [2], [5], agriculture [6], [7], and industrial drying [8], [9]. Up to date, most of the RH sensors are either electrical or optical. Compared to electrical RH sensors, optical RH sensors have the advantages of fast response time, lightweight, and small footprint. Particularly, optical fiber RH sensors hold an unparalleled advantage over electrical sensors when it comes to signal transmission. Unlike electrical sensors that require complex devices for signal transmission [10], optical fiber sensors allow the signal to be naturally coupled to the output fiber and transmit with low loss and high fidelity without being interfered by the environment. This advantage is especially desired in many RH sensing applications, which often require remote readout in a hazard or electromagnetic-interference environment [10], [11]. Several types of optical fiber RH sensors have been previously demonstrated, including the optical fiber Fabry-Pérot RH sensors [12], tapered optical fiber RH sensors [13], and optical fiber RH sensors based on the surface plasmon resonance [14]. In recent years, a new type of optical fiber sensor based on the lossy mode resonance (LMR) has been proposed [15]–[17]. It has been demonstrated that the LMR fiber sensors have an ultrahigh sensitivity [18] in environmental refractive index sensing. They have found an array of applications ranging from biosensing [19] to environmental monitoring [20].

Mainly two types of LMR optical fiber sensors have been demonstrated: cylindrically symmetric LMR fiber sensors with the sensing element coated on the cylindrical fiber core [21], [22], and asymmetric D-shaped LMR fiber sensors with the sensing elements coated on the side-polished surfaces of the D-shape fibers [16], [17]. Compared to the cylindrically symmetric LMR fiber sensors, the D-shaped LMR fiber sensors have several advantages. Firstly, cylindrically symmetric LMR fiber sensors are less flexible. This is because almost all the cylindrically symmetric LMR fiber sensors are based on multi-mode fibers with cores larger than 200 $\mu m$ [16], [17], while the D-shaped LMR fiber sensors can be made from the more flexible single-mode fibers with a height of less than 70 $\mu m$ at the D-shaped sensing part. Secondly, the single-mode fibers have lower transmission loss and allow better polarization control of the input light. The polarization control is important because the LMR resonance supports both the TE and the TM modes. These two modes often have different but very close resonance wavelengths, so their overlapping may cause distortion of the resonance shape [17]. Furthermore, these two modes have different sensitivity to environmental changes, which may result in distorted sensor responses. In fact, the cylindrically symmetric LMR fiber sensors not only cannot well control the input light polarization, but they also intrinsically cannot distinguish the TE and TM LMR modes [17] because of their symmetrical structure. As a result, the mechanically more flexible D-shaped LMR single-mode fiber sensors that have a smaller dimension and allow the isolation of the TE and TM LMR modes are always desired.

Another aspect of the sensor design is the sensor's form factor. To the best of our knowledge, only LMR fiber sensors with the in-line form factor have been demonstrated [15], [16], [18], [21], [23], [24]. The in-line form factor requires two separate ports for optical input and output, which can be challenging to realize in spatially confined applications. Adding to this challenge is the requirement of keeping the middle region of the D-shaped fiber aligned along a straight line to prevent it from breaking due to bending or stretching. Such strict requirements of the fiber alignment can be difficult to satisfy in an environment with complex geometry. These challenges made the in-line LMR fiber sensors cumbersome to implement in real-world applications. One LMR fiber sensor design that overcomes the above-mentioned disadvantages is the fiber-tip sensor design. Different from the in-line sensor design, LMR fiber sensors with the fiber-tip form factor require only a single port serving both as the optical input and output. Furthermore, the free-standing fiber tip is much less prone to breakages due to bending or stretching. These advantages endow the fiber-tip LMR sensor with high implementation flexibility and durability.

The dynamic response of LMR fibers is important for applications that require real-time monitoring of rapidly changing RH signals. Previous studies have characterized the LMR fiber sensors' dynamic responses using the imprecise RH changes induced by the human breath [25] and slowly changing RH over the time span of hours [26], [27]. However, LMR fiber sensors' dynamic response to rapidly changing RH using a controlled environmental chamber is yet to be characterized.

In this work, we developed a D-shaped fiber-tip LMR sensor for RH sensing applications. The D-shaped fiber tip used in this work was fabricated using an in-house fiber polishing setup improved from a similar setup in the previous work [28]. The improved setup enables the fabrication of D-shaped fiber tips in a highly repeatable and straightforward manner. The LMR spectrum of the fiber-tip sensor was investigated both experimentally and using numerical simulations. The LMR fiber sensor's responses to RH changes were experimentally characterized using a custom-built RH chamber. The characterization results show that the LMR fiber sensor has a wide linear response ranging from ~6% to ~75% RH and a resolution of around 4% RH. With the custom-built RH chamber, we characterized the LMR fiber-tip sensor's dynamic response to rapid and repeated RH changes between ~5.0% and ~86.0%. The results show that our sensor's capability of measuring dynamic RH changes is comparable to that of the electrical reference sensor. Thanks to the fiber-tip form factor, the developed LMR fiber-tip RH sensor enables RH sensing in hazardous and spatially constrained environments, which is challenging for electrical sensors and in-line LMR fiber sensors. Our fiber-tip sensor will particularly benefit industrial drying applications where accurate and real-time RH measurements can improve drying efficiency and prevent unnecessary energy waste. Furthermore, the D-shaped fiber tip form factor not only benefits LMR RH sensors but also other types of LMR fiber sensors, providing increased flexibility and expanding their applications.

## II. Working Principle and Fabrication of the Sensor

### A. Sensor Working Principle

Our D-shaped fiber-tip LMR RH sensor has a straightforward design, as shown in Figure 1. (a). The side-polished fiber tip, where the LMR sensing film is coated, is naturally connected to a regular single-mode fiber for efficient optical input and readout. The fiber tip has a D-shaped cross-section, as shown in Figure 1. (b). A thin layer of $SnO_2$ film is sputter coated onto the flat polished surface of the fiber tip, serving as the LMR sensing layer. When the input light from the regular single-mode fiber arrives at the D-shaped fiber tip, depending on its wavelength, the light may or may not be efficiently coupled into the $SnO_2$ film. Since the $SnO_2$ film has a complex refractive index with a non-zero imaginary part, the light at the LMR wavelength will couple into the $SnO_2$ film and experience an optical loss [16], while the light outside the LMR wavelength will propagate in the fundamental core mode with only negligible optical loss. As a result, if broadband light is input into the LMR fiber tip and reflected at the fiber end face, there will be a dip at the LMR wavelength in the optical spectrum, hence the name lossy mode resonance.

Two critical geometric parameters of our fiber-tip LMR sensor are the $SnO_2$ film thickness and the residual cladding thickness [29]. The $SnO_2$ film thickness determines the resonance wavelength of the LMR, while the residual cladding thickness determines the depth or the visibility of the LMR spectrum. To allow the LMR to be efficiently excited within the wavelength range of our light source, these thickness values were determined based on numerical simulations and experimental characterizations discussed in section III. Figure 1. (c) and (d) show the interaction between the coated $SnO_2$ film and the water molecules, which illustrates how the fiber-tip LMR sensor senses the environmental RH change. As shown in Figure 1. (d), right after the $SnO_2$ film was coated onto the fiber tip, water molecules were chemisorbed by the $SnO_2$ and formed a layer of hydroxyl group on the $SnO_2$ film surface [30]. Since both the hydroxyl group and the water molecules are dipolar, more water molecules can be physisorbed onto the hydroxyl layer of the $SnO_2$ and the already physisorbed water molecule layer [31], as shown in Figure 1. (c) and (d). Higher relative humidity will result in a thicker physisorbed water molecule layer, which will increase the environmental refractive index seen by the light propagating in the fiber tip. Such change will cause a redshift of the LMR spectrum and vice versa. Furthermore, while the water chemisorption process of forming the hydroxyl group on the $SnO_2$ film is not reversible, the water physisorption process is reversible [26]. These properties make the $SnO_2$ film suitable for RH sensing applications.

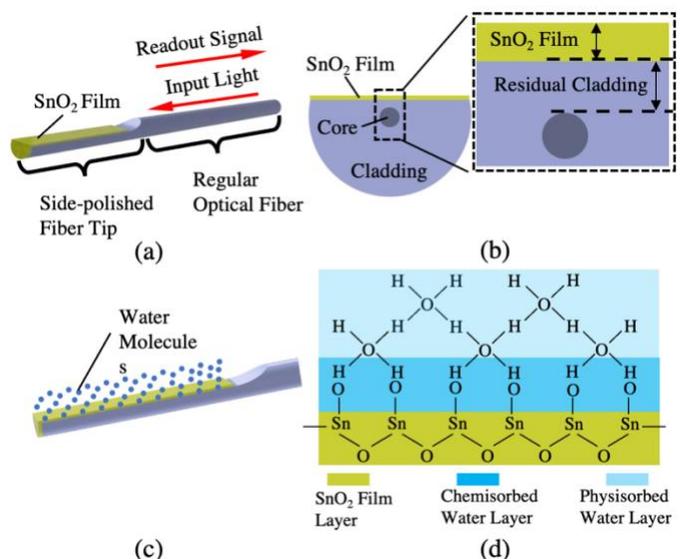

Figure 1. (a) Schematic of the D-shaped fiber-tip LMR sensor. (b) Cross-section view of the D-shaped fiber tip with the $SnO_2$ coating. The zoom-in view shows the $SnO_2$ film and the residual cladding thickness. (c) and (d) illustrates the interaction between the $SnO_2$ film and the environmental water molecules. In (d), the yellow layer represents the $SnO_2$ film, the darker blue layer represents the chemisorbed water molecules, and the light blue layer represents the physisorbed water molecules.

### B. Sensor Fabrication

As mentioned before, D-shaped LMR fiber sensors with a fiber-tip form factor have the advantage of being flexible in implementations. However, up to date, most of the existing fabrication methods for D-shaped LMR fiber sensors are based on in-line D-shaped optical fibers. Here, we present a highly reproducible fabrication method of D-shaped LMR sensors with a fiber tip form factor. The photo and schematic of the fabrication setup are shown in Figure 2. (a) and (b). The fabrication process of the D-shaped fiber-tip LMR sensor is shown in Figure 2. (c) and can be summarized into three steps. First, a single-mode optical fiber (SMF-28, Corning) is side-polished into an in-line D-shaped fiber. Then, the polished D-shaped fiber is split into two D-shaped fiber tips in one step. Finally, a $SnO_2$ film was sputter coated onto the D-shaped fiber tips to endow them with the LMR-based RH sensing capability. Details of the fabrication setup and process are discussed in the following paragraph.

To prepare for the polishing, as shown in Figure 2. (a), a single-mode optical fiber with stripped plastic buffer was partially wrapped around a cylindrical polishing rod. The polishing rod consists of two parts, a 1-inch diameter aluminum rod and a polishing paper (FiberMet Abrasive Discs, BUEHLER) with a roughness of 12 μm. The polishing paper has glue on one side, which allows it to attach to the aluminum rod surface. The polishing paper was cut into a rectangular shape. The length of the cutout polishing paper is designed to match the circumference of the aluminum cylinder to minimize the gap between the two edges of the polishing paper after wrapping it around the aluminum rod. This creates a smooth polishing surface, which is critical for the fabrication. The polishing rod was connected to a DC motor. During the polishing, the motor drives the polishing rod to rotate at 60 rpm. The two free ends of the fiber were clamped onto two optical fiber clamps (SM1F1-250, ThorLabs), which were mounted on two motorized stages (MTS50-Z8, ThorLabs). Different from the optical fiber side-polishing setup in the previous work [17], our fiber was also in contact with two supporting rods, as shown in Figure 2. (a) and (b). The supporting rods have two functions. Firstly, they reduce the vibration of the fiber during the polishing by introducing frictional damping between the fiber and the supporting rod. The reduced vibration of the fiber during the polishing resulted in a smoother polished fiber surface. Secondly, they ensure the fiber is bent into a proper curvature to partially wrap around the polishing rod. The polishing length is determined by the length of the fiber that is in contact with the polishing rod. In our setup, the polishing length is around 17-18 mm. The polishing of the fiber relies on the friction force between the fiber's surface and the polishing paper. This friction force is determined by both the roughness of the polishing paper and the normal contact force between the fiber and the polishing rod. An overly large friction force will cause the polished fiber surface to be too rough to be used for LMR sensing. While a smaller friction force creates a smoother polished surface, it may cause the polishing time to be impractically long. In our experiment, an optimum friction force that balances the polished surface roughness and the polishing time was achieved by adjusting the relative position between the polishing rod and the supporting rods and a pre-applied tension along the fiber. During the polishing, a 1310 nm laser was input into the fiber, and its transmission was used to monitor the residual cladding thickness. At the beginning of the polishing, because the cladding being polished was far from the light-guiding fiber core, the fiber's transmission was not affected. However, as the residual cladding thickness decreased to less than 6 μm, the evanescent field of the guided light in the fiber core started to see the surface that was being polished. As a result, part of the guided light inside the fiber was scattered by the polishing surface, and the optical transmission started to decrease rapidly. Based on our experimental results, stopping the polishing at an optical transmission loss of around 0.05 dB will result in a residual cladding thickness of around 3.0-4.5 μm. After the polishing, both the polishing rod and the supporting rods were moved away from the fabricated in-line D-shaped fiber. Finally, to fabricate D-shaped fiber tips, we moved the two fiber ends clamped to the motorized stages rapidly away from each other (0.1 mm/s each stage) to stretch and break the polished fiber into two D-shaped fiber tips. The split fiber tips had flat end faces, as shown in Figure 2. (d). The flat end face of the D-shaped fiber tip can efficiently reflect the input light, which allows the LMR spectrum to be read out from the same input fiber port.

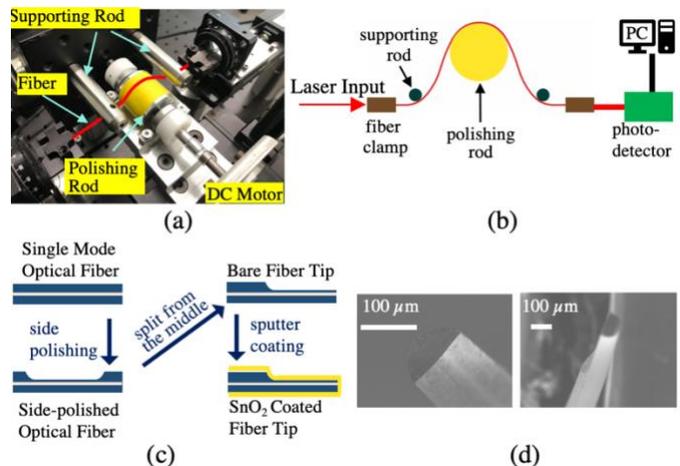

Figure 2. (a) and (b) show the photo and the schematic of the optical fiber polishing setup. The optical fiber is highlighted by a red line in (a). (b) shows the fabrication steps of the D-shaped fiber-tip LMR sensor. The sequence of the steps is indicated by the arrows. (d) shows the SEM images of the polished D-shaped fiber tip.

### III. SENSOR CHARACTERIZATION

We have conducted a detailed characterization of the fiber-tip LMR sensor. The characterization results are presented in the following sections. In section A, the LMR spectrum of the fiber-tip sensor was characterized both experimentally and using numerical simulations. In sections B and C, the fiber-tip LMR sensor's response to RH changes, including sensitivity, resolution, response time, and reversibility, were characterized.

#### A. Characterization of the LMR Spectrum

To better understand the LMR supported by the $SnO_2$ film coated on the fiber tip, we studied the LMR spectrum using the finite element method (FEM) and compared the simulation results with the experimentally measured LMR spectrum. As shown in Figure 3, the FEM simulation results agreed well with our experimental measurements. The experimental setup used for the characterization is shown in Figure 3. (a). This same setup was used for all other sensor characterizations presented later. A broadband superluminescent diode SLD with a wavelength range of 1200 – 1600 nm was used as the light source. The polarization of the light was controlled using an in-line polarization controller (FPC562, Thorlabs). The polarization of the input light was adjusted for the best visibility of the LMR signal. Based on our numerical simulations, which will be detailed later, the polarization that produces the optimum LMR visibility has an electric field polarized parallel to the film along the x-axis, as indicated in Figure 3. (b). An optical circulator was used to couple light into the fiber-tip LMR sensor and route the reflected optical signal from the fiber tip into the optical spectrometer (Sol 1.7, B&W Tek). The fiber-tip LMR sensor used in the experiments has a $SnO_2$ film thickness of 190 nm and a residual cladding thickness of 3.5 μm. As mentioned above, the thickness of the $SnO_2$ film, which was the main factor that determines the LMR wavelength, was designed to allow the LMR to sit within the wavelength range of our SLD light source. The residual cladding thickness, which was the main factor that determines the LMR's coupling efficiency, was designed to allow an optimum light coupling into the $SnO_2$ film. In our experiments, we found that a residual cladding thickness thinner than 2 $\mu m$ resulted in large optical loss over the whole wavelength range and low signal-to-noise ratio; and a residual cladding thickness thicker than 4 $\mu m$ resulted in low excitation efficiency of the LMR mode, which resulted in low resonance visibility. A typical normalized LMR spectrum from the fiber-tip sensor was shown in Figure 3. (a). The spectrum was normalized using the reflected spectrum from a $SnO_2$-coated cleaved fiber tip of a non-polished regular SMF-28 fiber. The resonance wavelength of the experimentally measured LMR spectrum was located at around 1338 nm. It is noted that the periodic oscillation signal at the right arm of the LMR spectrum (between ~1400 nm to ~1500 nm) was due to the interference at the non-ideal fiber coupler. Because the amplitude of this interference-induced oscillation was much smaller than the depth of the lossy mode resonance, its influence on the sensor's performance was negligible.

core mode and the LMR mode is satisfied, so that the core mode can most efficiently excite the lossy mode resonance of the $SnO_2$ film, which will result in a maximum transmission loss. To confirm this result, we have also calculated the normalized transmission of the LMR fiber tip based on the FEM simulations as shown in the lower plot of Figure 3. (b). In the calculation, we assumed that the optical loss of the core mode comes only from the optical absorption in the $SnO_2$ film. The normalized transmission of the LMR fiber tip was calculated using the following equations:

$$T = 1 - e^{-\alpha * L} \quad (1)$$
$$\alpha = \frac{2\pi * Im[n_{eff}]}{\lambda} \quad (2)$$

In the above equations, $T$ is the normalized transmission, $\alpha$ and $n_{eff}$ are the propagation loss coefficient and the complex effective mode index of the guided LMR mode respectively, $L$ is the length of the $SnO_2$ film coating, $\lambda$ is the input light wavelength. In our calculations, $L$ was set to 10 mm based on the experimental data, $n_{eff}$ values at different wavelengths were obtained from the FEM simulations. The normalized plot confirms that the lowest transmission happens at the wavelength where the $TE_{11}$ and $HE_{11}$ have the same real effective mode index, as indicated by the black dashed line in Figure 3. (b). The insets in the normalized transmission plot were the electric field distributions of the D-shaped fiber when excited by an x-polarized $HE_{11}$ mode. These plots show that at the LMR wavelength, the $TE_{11}$ and $HE_{11}$ modes have significant overlapping, which enables the efficient excitation of the LMR mode. It is noted that, while both the x-polarized and y-polarized $HE_{11}$ mode can excite the LMR in the $SnO_2$ film, our simulations showed that within the wavelength range of 1200 nm to 1600 nm, only the $TE_{11}$ mode LMR will be excited by the x-polarized $HE_{11}$ mode. These simulation results matched our experimental characterizations, where only one polarization mode of LMR was found between 1200 nm to 1600 nm.

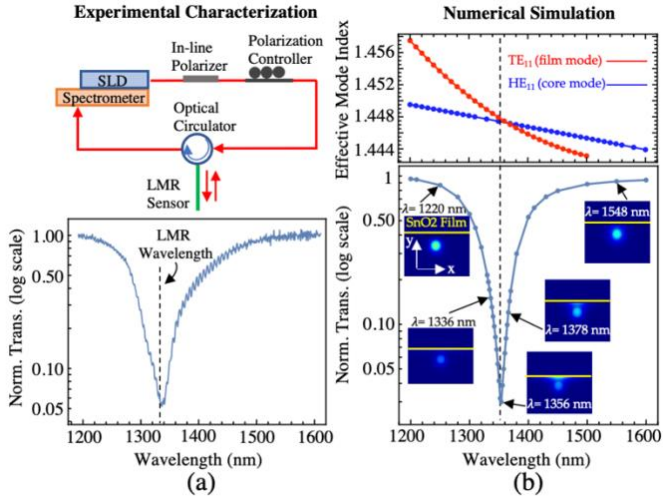

Figure 3. (a) Optical characterization setup for the fiber-tip sensor. The bottom plot in (a) shows a typical LMR spectrum from the fiber sensor. (b) Numerical study of the LMR. The top plot shows the dispersion curve of the $TE_{11}$ and $HE_{11}$ mode. The y-axis is the real part of the effective mode index, and the x-axis is the wavelength. The top and bottom plots have identical wavelength ticks. The bottom plot shows the calculated LMR spectrum based on the FEM simulation. The insets show the simulation results of the electric field of the D-shaped fiber tip at different wavelengths. The polarization of the electric field is along the x-axis of the inset.

The numerical simulation results of the fiber-tip LMR sensor were shown in Figure 3. (b). The simulations were conducted in a commercial FEM software (COMSOL). In the simulations, the $SnO_2$ film thickness and the residual cladding thickness were chosen to match that used in the experimental characterization mentioned above. The refractive index of the $SnO_2$ film was obtained from ellipsometry measurements. The refractive index of the glass fiber's core and cladding were calculated using the well-known Sellmeier relation [18].

The upper plot of Figure 3. (b) shows the real part of the effective mode index of the $TE_{11}$ film mode and the $HE_{11}$ core mode at different wavelengths. Here, the effective mode index is defined as the propagation constant of the mode divided by the free-space wavenumber. The intersection of the $TE_{11}$ and $HE_{11}$ curves in Figure 3. (b) corresponds to the wavelength where the real parts of their effective mode indices are identical. At this wavelength, the phase matching condition between the

### B. *RH Sensitivity and Resolution of the Sensor*

In this section, we detailed the experimental characterizations of the fiber-tip LMR sensor's RH responses. In the experiments, the fiber-tip sensor and a commercially available electrical RH sensor [Omega, RH-USB] were enclosed in a customized climate chamber. The climate chamber has two air inlets and one air outlet for humidity control. The two inlets were connected to an RH>90% humid air input and an RH<3% dry air input, respectively. The outlet of the chamber was connected to the room environment. By adjusting the flow rate ratio of the humid and dry air input, the RH value within the chamber can be changed between ~4% and ~85%.

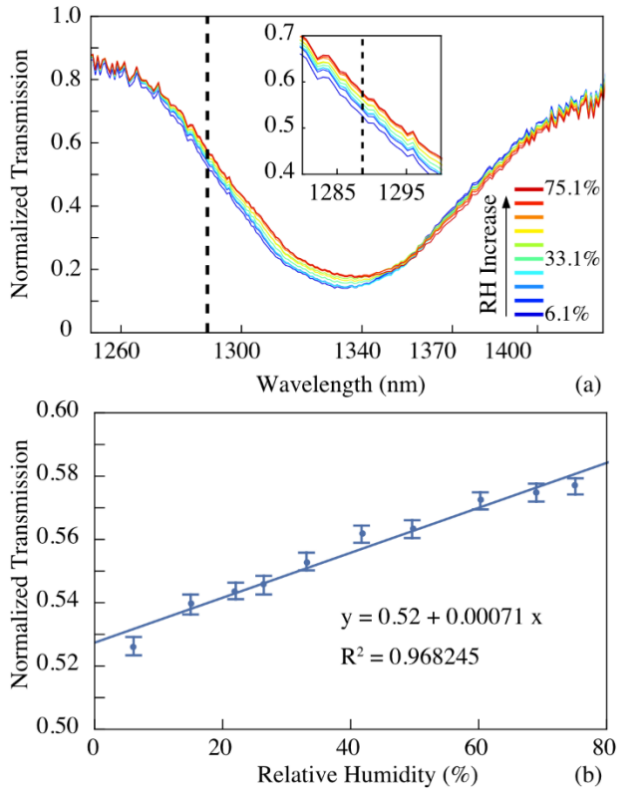

Figure 4. (a) LMR spectra experimentally measured at RH between 6.1% and 75.0%. Spectrum measured at different RH was plotted in different colors as indicated in the plot legend. At each RH, 244 spectra were measured, and the averaged spectrum was plotted. The inset shows the zoom-in of the shoulder of the LMR spectra. The black dashed line indicates the LMR shoulder intensities used for the characterizations. (b) LMR shoulder intensities at different RH values. The error bar was the standard deviation calculated from the 244 spectra. The inset equations show the linear fitting results and the R-squared value of the fitting.

To characterize our sensor's sensitivity and resolution, the fiber-tip LMR sensor's responses at RH values between 6.1% and 75.0% were experimentally measured, as shown in Figure 4. (a). The measurement results show that the increase of the RH values caused the LMR spectra to red shift. The red shift, as shown in the zoom-in plot in Figure 4. (a)'s inset, corresponded to an increase of the optical transmissions at the shoulder of the LMR spectrum. In Figure 4. (b), the optical transmissions at different RH values at the LMR's shoulder (indicated by the dashed black line) were plotted. The plot shows that the optical transmission at the shoulder of the LMR spectra changes linearly with the RH value, which makes it an ideal signal for RH sensing. The fitting in Figure 4. (b) shows that our LMR fiber-tip sensor has a sensitivity of around 0.00071 RH$^{-1}$. The sensor's RH resolution $r_{RH}$ was calculated using $r_{RH} = \sigma/S$, where $\sigma$ is the stability of the sensor and S is the sensitivity of the sensor [32]. The stability $\sigma$ was defined as the standard deviation of the optical transmission from multiple measurements. Specifically, $\sigma$ was calculated from 244 measurements repeated in an environment with an RH of around 22%. With the experimentally acquired sensitivity and stability, the resolution of the RH sensor is calculated to be as good as 4%. It is noted that several factors that are not fundamental to the LMR fiber sensor have negatively influenced the sensors' resolution characterization results. Such factors include the noises from the laser source and the spectrometer and polarization perturbation of the input light due to the bending and vibration of the fiber. Particularly for the polarization perturbation, since the TE and TM polarized LMR modes have very different spectrum shapes, such a perturbation will significantly reduce the sensor's stability $\sigma$, and hence reduce the sensor's resolution. Furthermore, it is noted that the RH inside the customized climate chamber during each measurement has certain fluctuation. Such fluctuation also had a negative impact on the sensor's resolution characterization results. Experiments are underway using an improved climate box that can provide a more stable RH environment for sensor characterizations. We are also developing a D-shaped fiber sensor fabricated from a polarization-maintaining fiber (results not shown), which can prevent light depolarization. With these improvements, we expect our fiber-tip LMR sensor to achieve an RH resolution much better than 4% in the future.

### C. Sensor Response to Dynamic RH

Our LMR fiber RH sensor's response time and reversibility in an environment with rapidly changing RH were characterized. Here, the response time is characterized by the time needed by the sensor to provide a stable readout when the RH of the environmental box is changed rapidly. The reversibility is determined by the sensor's capability to reverse to its initial state when the environmental humidity repeatedly changes between low and high. The fast response time is important when the sensor is responsible for measuring a rapidly changing dynamic RH. Reversibility is necessary when a single sensor is responsible for monitoring either a process with multiple cycles or multiple processes in succession.

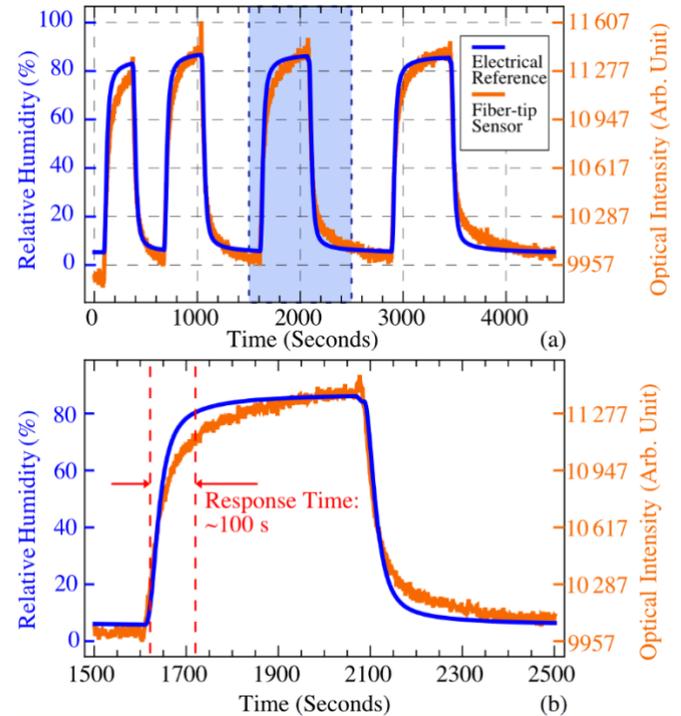

Figure 5. (a) and (b) show the characterization of the sensor's response time and reversibility. (a) shows the response of the sensor to four RH cycles. In each cycle, the RH changes from ~5.0% to ~86.0%. The blue line shows the RH measured by the electrical reference sensor. The orange line shows the optical signal measured by the fiber-tip sensor. (b) shows the sensor's optical response in a single RH cycle (blue highlighted cycle in (a)). The two red dashed lines indicate the duration before the LMR signal stabilized. (b) has the same color codes as (a).

Experimentally, we characterized the sensor's response time and reversibility by cycling the RH of the climate chamber between ~5.0% and ~86.0% for four times. Both the optical responses from our fiber-tip sensor and the RH values read out from the reference electrical RH sensor [Omega, RH-USB] were recorded at a frequency of 1 Hz, as shown in Figure 5. (a). The optical intensities were measured at the same shoulder position of the LMR spectra as in Figure 4. The overlapping of the optical sensor signal and the reference electrical signal showed that the optical fiber-tip sensor was able to faithfully characterize the fluctuating RH in all cycles. Furthermore, as shown in Figure 5. (a) and (b), the LMR signal was able to quickly reverse at the end of each cycle when the RH value falls from ~86.0% back to ~5.0%. These results demonstrated that the fiber-tip LMR sensor had a decent reversibility and response time comparable to those of the electrical reference sensor. To further characterize the sensor's response time, a single response cycle was zoomed in and studied as shown in Figure 5. (b). It shows that in response to the rapid RH change, the fiber-tip sensor's LMR signal stabilized in around 100 seconds. While this is slightly slower than that of the electrical reference sensor, it is sufficient for many drying applications. This is because the RH change in many drying applications is relatively slow, a drying process of around 80% RH drop can often be longer than 100 seconds [33], [34]. More importantly, the optical fiber-tip sensor can enable RH measurements in applications where the complex environment forbids the use of electrical RH sensors, such as in microwave drying applications. In these scenarios, albeit with a slightly slower response time, our optical fiber-tip LMR sensor is still beneficial because it can provide valuable RH measurements that are not obtainable by the electrical sensors.

## IV. CONCLUSIONS

In conclusion, we have developed a fiber-tip LMR RH sensor that enables real-time RH monitoring in harsh environments. The developed sensor is the first experimental demonstration of an LMR fiber sensor with a fiber-tip form factor. We have developed a customized optical fiber side polishing system that allows highly repeatable D-shaped fiber tip fabrications. The LMR spectrum of the fiber-tip sensor was studied both experimentally and numerically, whose results explained the fundamental mechanism of the excitation of the LMR on the $SnO_2$ film. The LMR fiber-tip sensor's capability of sensing environmental RH changes through the chemisorption and physisorption of water molecules was experimentally characterized. The characterization results have shown that the LMR fiber-tip sensor has a linear response to RH changes in the range of 6.1% and 75.0% RH, with a sensitivity of 0.00071 $RH^{-1}$ and a resolution better than 4% RH. The LMR fiber-tip sensor's responses to dynamic RH that rapidly changes between 5.2% and 86.7% were experimentally characterized. These characterization results showed that the developed fiber-tip LMR RH sensor meets the urgent need for real-time RH monitoring in harsh and critical environments. Beyond RH sensing, our innovative D-shaped fiber-tip design is also applicable to LMR fiber sensors in general and may benefit a broad range of applications.

## V. ACKNOWLEDGMENT

This study was partially supported by the US Department of Energy (#DE-FOA-0001980), the Center for Advanced Research in Drying (CARD), and the Massachusetts Clean Energy Center (MassCEC). CARD is a US National Science Foundation Industry University Cooperative Research Center. CARD is located at Worcester Polytechnic Institute, with the University of Illinois at Urbana-Champaign as the co-site.

## VI. REFERENCE:

[1] H. Yan, D. Han, M. Li, and B. Lin, "Relative humidity sensor based on surface plasmon resonance of D-shaped fiber with polyvinyl alcohol embedding Au grating," *J Nanophotonics*, vol. 11, no. 1, p. 016008, Feb. 2017, doi: 10.1117/1.JNP.11.016008.

[2] F. Bibi, C. Guillaume, A. Vena, N. Gontard, and B. Sorli, "Wheat gluten, a bio-polymer layer to monitor relative humidity in food packaging: Electric and dielectric characterization," *Sens Actuators A Phys*, vol. 247, pp. 355–367, Aug. 2016, doi: 10.1016/J.SNA.2016.06.017.

[3] A. Vena, B. Sorli, Y. Belaizi, B. Saggin, and J. Podlecki, "An Inkjet Printed RFID-Enabled Humidity Sensor on Paper Based on Biopolymer," *2018 2nd URSI Atlantic Radio Science Meeting, AT-RASC 2018*, Sep. 2018, doi: 10.23919/URSI-AT-RASC.2018.8471445.

[4] S. Hajian *et al.*, "Development of a Fluorinated Graphene-Based Flexible Humidity Sensor," *FLEPS 2019 - IEEE International Conference on Flexible and Printable Sensors and Systems, Proceedings*, Jul. 2019, doi: 10.1109/FLEPS.2019.8792254.

[5] Y. Wang, Q. Huang, W. Zhu, and M. Yang, "Simultaneous Measurement of Temperature and Relative Humidity Based on FBG and FP Interferometer," *IEEE Photonics Technology Letters*, vol. 30, no. 9, pp. 833–836, May 2018, doi: 10.1109/LPT.2018.2818744.

[6] L. Lan, X. Le, H. Dong, J. Xie, Y. Ying, and J. Ping, "One-step and large-scale fabrication of flexible and wearable humidity sensor based on laser-induced graphene for real-time tracking of plant transpiration at bio-interface," *Biosens Bioelectron*, vol. 165, p. 112360, Oct. 2020, doi: 10.1016/J.BIOS.2020.112360.

[7] T. Cao-Hoang and C. N. Duy, "Environment monitoring system for agricultural application based on wireless sensor network," *7th International Conference on Information Science and Technology, ICIST 2017 - Proceedings*, pp. 99–102, May 2017, doi: 10.1109/ICIST.2017.7926499.

[8] J. Li *et al.*, "Fuzzy logic control of relative humidity in microwave drying of hawthorn," *J Food Eng*, vol.

- [9] U. Akyol, A. E. Akan, and A. Durak, "Simulation and thermodynamic analysis of a hot-air textile drying process," *http://dx.doi.org/10.1080/00405000.2014.916062*, vol. 106, no. 3, pp. 260–274, Mar. 2015, doi: 10.1080/00405000.2014.916062.
- [10] Y. Zhao, R. jie Tong, M. Q. Chen, and F. Xia, "Relative humidity sensor based on hollow core fiber filled with GQDs-PVA," *Sens Actuators B Chem*, vol. 284, pp. 96–102, Apr. 2019, doi: 10.1016/J.SNB.2018.12.130.
- [11] L. Xia, L. Li, W. Li, T. Kou, and D. Liu, "Novel optical fiber humidity sensor based on a no-core fiber structure," *Sens Actuators A Phys*, vol. 190, pp. 1–5, Feb. 2013, doi: 10.1016/J.SNA.2012.10.041.
- [12] A. Vaz, N. Barroca, M. Ribeiro, A. Pereira, and O. Frazao, "Optical Fiber Humidity Sensor Based on Polyvinylidene Fluoride Fabry-Perot," *IEEE Photonics Technology Letters*, vol. 31, no. 7, pp. 549–552, Apr. 2019, doi: 10.1109/LPT.2019.2901571.
- [13] M. Batumalay *et al.*, "Tapered plastic optical fiber coated with ZnO nanostructures for the measurement of uric acid concentrations and changes in relative humidity," *Sens Actuators A Phys*, vol. 210, pp. 190–196, Apr. 2014, doi: 10.1016/J.SNA.2014.01.035.
- [14] J. Wang, "Surface plasmon resonance humidity sensor based on twisted long period fiber grating coated with tungsten disulfide film," *Optik (Stuttg)*, vol. 236, p. 166616, Jun. 2021, doi: 10.1016/J.IJLEO.2021.166616.
- [15] I. Del Villar *et al.*, "Optical sensors based on lossy-mode resonances," *Sens Actuators B Chem*, vol. 240, pp. 174–185, Mar. 2017, doi: 10.1016/J.SNB.2016.08.126.
- [16] A. Ozcariz, C. R. Zamarreño, P. Zubiate, and F. J. Arregui, "Is there a frontier in sensitivity with Lossy mode resonance (LMR) based refractometers?," *Scientific Reports 2017 7:1*, vol. 7, no. 1, pp. 1–7, Aug. 2017, doi: 10.1038/s41598-017-11145-9.
- [17] A. Ozcariz, C. Ruiz-Zamarreño, and F. J. Arregui, "A Comprehensive Review: Materials for the Fabrication of Optical Fiber Refractometers Based on Lossy Mode Resonance," *Sensors 2020, Vol. 20, Page 1972*, vol. 20, no. 7, p. 1972, Apr. 2020, doi: 10.3390/S20071972.
- [18] F. J. Arregui, I. Del Villar, C. R. Zamarreño, P. Zubiate, and I. R. Matias, "Giant sensitivity of optical fiber sensors by means of lossy mode resonance," *Sens Actuators B Chem*, vol. 232, pp. 660–665, Sep. 2016, doi: 10.1016/J.SNB.2016.04.015.
- [19] F. Chiavaioli *et al.*, "Femtomolar Detection by Nanocoated Fiber Label-Free Biosensors," *ACS Sens*, vol. 3, no. 5, pp. 936–943, May 2018, doi: 10.1021/ACSSENSORS.7B00918.
- [20] C. Elosua *et al.*, "Volatile organic compounds optical fiber sensor based on lossy mode resonances," *Sens Actuators B Chem*, vol. 173, pp. 523–529, Oct. 2012, doi: 10.1016/J.SNB.2012.07.048.
- [21] M. Smietana *et al.*, "Optical Monitoring of Electrochemical Processes With ITO-Based Lossy-Mode Resonance Optical Fiber Sensor Applied as an Electrode," *Journal of Lightwave Technology*, vol. 36, no. 4, pp. 954–960, Feb. 2018, doi: 10.1109/JLT.2018.2797083.
- [22] M. Hernaez, A. G. Mayes, and S. Melendi-Espina, "Graphene Oxide in Lossy Mode Resonance-Based Optical Fiber Sensors for Ethanol Detection," *Sensors 2018, Vol. 18, Page 58*, vol. 18, no. 1, p. 58, Dec. 2017, doi: 10.3390/S18010058.
- [23] C. R. Zamarreño, M. Hernaez, I. Del Villar, I. R. Matias, and F. J. Arregui, "Lossy mode resonance-based optical fiber humidity sensor," *Proceedings of IEEE Sensors*, pp. 234–237, 2011, doi: 10.1109/ICSENS.2011.6127124.
- [24] Q. Wang and W. M. Zhao, "A comprehensive review of lossy mode resonance-based fiber optic sensors," *Opt Lasers Eng*, vol. 100, pp. 47–60, Jan. 2018, doi: 10.1016/J.OPTLASENG.2017.07.009.
- [25] M. Hernaez, B. Acevedo, A. G. Mayes, and S. Melendi-Espina, "High-performance optical fiber humidity sensor based on lossy mode resonance using a nanostructured polyethylenimine and graphene oxide coating," *Sens Actuators B Chem*, vol. 286, pp. 408–414, May 2019, doi: 10.1016/J.SNB.2019.01.145.
- [26] J. Ascorbe, J. M. Corres, I. R. Matias, and F. J. Arregui, "High sensitivity humidity sensor based on cladding-etched optical fiber and lossy mode resonances," *Sens Actuators B Chem*, vol. 233, pp. 7–16, Oct. 2016, doi: 10.1016/J.SNB.2016.04.045.
- [27] C. R. Zamarreño, M. Hernaez, I. Del Villar, I. R. Matias, and F. J. Arregui, "Tunable humidity sensor based on ITO-coated optical fiber," *Sens Actuators B Chem*, vol. 146, no. 1, pp. 414–417, Apr. 2010, doi: 10.1016/J.SNB.2010.02.029.
- [28] S. Omar, B. Musa, Z. Jusoh, M. Ali, A. Jafry, and S. Harun, "Passively Q-switched Erbium-doped Fiber Laser using Tungsten Disulfide deposited D-shaped Fiber as Saturable Absorber," *IOP Conf Ser Mater Sci Eng*, vol. 854, no. 1, p. 012021, May 2020, doi: 10.1088/1757-899X/854/1/012021.
- [29] C. R. Zamarreño, F. J. Arregui, I. Del Villar, I. R. Matias, and P. Zubiate, "Optimization in nanocoated D-shaped optical fiber sensors," *Optics Express, Vol. 25, Issue 10, pp. 10743-10756*, vol. 25, no. 10, pp. 10743–10756, May 2017, doi: 10.1364/OE.25.010743.
- [30] B. C. Yadav, N. K. Pandey, A. K. Srivastava, and P. Sharma, "Optical humidity sensors based on titania films fabricated by sol–gel and thermal evaporation methods," *Meas Sci Technol*, vol. 18, no. 1, p. 260, Dec. 2006, doi: 10.1088/0957-0233/18/1/032.
- [31] W. M. Sears, "The effect of oxygen stoichiometry on the humidity sensing characteristics of bismuth iron molybdate," *Sens Actuators B Chem*, vol. 67, no. 1–2, pp. 161–172, Aug. 2000, doi: 10.1016/S0925-4005(00)00395-6.
- [32] C.-H. Huang, G.-L. Cheng, H.-C. Chui, H.-Y. Lin, and N.-K. Chen, "Tapered optical fiber sensor based on localized surface plasmon resonance," *Optics Express,*

310, p. 110706, Dec. 2021, doi: 10.1016/J.JFOODENG.2021.110706.

*Vol. 20, Issue 19, pp. 21693-21701*, vol. 20, no. 19, pp. 21693–21701, Sep. 2012, doi: 10.1364/OE.20.021693.

[33] Md. I. H. Khan, T. Farrell, S. A. Nagy, and M. A. Karim, "Fundamental Understanding of Cellular Water Transport Process in Bio-Food Material during Drying," *Scientific Reports 2018 8:1*, vol. 8, no. 1, pp. 1–12, Oct. 2018, doi: 10.1038/s41598-018-33159-7.

[34] S. Kiani, S. Minaei, and M. Ghasemi-Varnamkhasti, "Real-time aroma monitoring of mint (Mentha spicata L.) leaves during the drying process using electronic nose system," *Measurement*, vol. 124, pp. 447–452, Aug. 2018, doi: 10.1016/J.MEASUREMENT.2018.03.033.